\documentstyle[twocolumn,epsfig,aps]{revtex}
\draft
\begin{document}
\title{Atom-photon entanglement generation and distribution}
\author{B. Sun, M. S. Chapman, and L. You}
\address{School of Physics, Georgia Institute of Technology,
Atlanta, GA 30332, USA}
\date{\today}
\maketitle
\begin{abstract}
We extend an earlier model by Law {\it et al.} \cite{law}
for a cavity QED based single-photon-gun
to atom-photon entanglement generation and distribution.
We illuminate the importance of a small critical atom
number on the fidelity of the proposed operation in
the strong coupling limit. Our result points to a
promisingly high purity and efficiency using currently
available cavity QED parameters, and
sheds new light on constructing quantum computing
and communication devices with trapped
atoms and high Q optical cavities.
\end{abstract}


Entanglement lies at the heart of quantum information and
computing science \cite{ekert,caves}---it is responsible both for
the mysteries of quantum cryptography \cite{qc1}
and teleportation \cite{tele}
as well as for the exponential speed-up promised by Shor's factoring
algorithm \cite{shor}. It is widely believed that the progress of
Quantum Information Science, on the experimental front,
will track closely the
progress of entanglement generation in the laboratory. Until
recently \cite{nist,haroche}, most experimental realizations of
entanglement came almost exclusively from photon down-conversion
using nonlinear crystals where a single pump photon spontaneously
converts into two correlated photons satisfying energy and
momentum conservation \cite{aspect,ou}. Although the individual
polarization states of photons are easily controlled, and their
quantum coherence can be preserved over many kilometers of
an optical fiber \cite{com2}, photons cannot be stored for long,
and manipulations of collective entangled state present
considerable difficulties even when photons are confined inside the same cavity.

The creation of long lived entangled pairs with
material particles (atoms and ions), on the other hand, is a relatively
recent pursuit \cite{cirac5,zo2,zo3}, spurred on in
large part by developments in quantum logic and computing.
These experimental efforts have been very
successful and are highlighted by the demonstration of a
4-ion entangled state \cite{sackett}, using a proposal
with trapped ions due to Molmer {\it et al.} \cite{mol}. However,
the scaling of this technology to larger numbers of qubits ($>10$)
is less certain, and it is unlikely that quantum information
stored exclusively in material particles will ever be effectively
distributed to remote locations as required for most quantum
communication and distributed computing applications.

Given the current state of affairs, there is a pressing need for
systems capable of integrating the relative strengths of material
particle-based entanglement and photon-based entanglement, wherein
the former provides reliable quantum information storage and local
entanglement capabilities, the latter provides quantum
communication capabilities over long distances. It is important
to develop capabilities for reliably converting and transferring
quantum information between material and photonic qubits.

In this article, we develop a system composed of a single trapped
atom inside a high Q optical cavity for deterministic
generation of atom-photon entanglement and its subsequent
distribution via the well directed photon from a high Q optical cavity.
For a large class of quantum communication protocols
(including cryptography protocols, teleportation, entanglement
purification, etc.), one always begins with the following statement:
`{\it Imagine that Alice has an entangled pair of particles, and
she sends one particle to Bob}...'. Our proposed system,
if implemented properly, will supply such a device.

\begin{figure}[h]
\includegraphics[width=3.in]{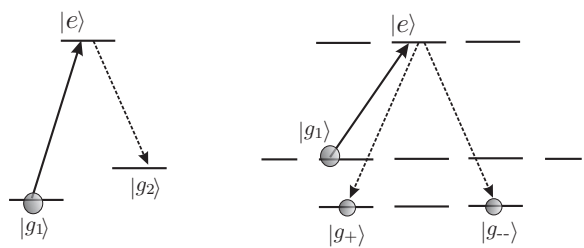}
\caption{Atom-photon entanglement illustrations.}
\label{fig1}
\end{figure}

The physical model for our system represents a direct extension of an earlier
proposal by Law {\it et al.} \cite{law} for a deterministic
single photon (or `Fock states') source \cite{sp},
an indispensable device for some quantum cryptographic
applications \cite{qc1}. Although significant
progress have already been made along the direction
of a deterministic single photon source \cite{sp},
to date, most experiments still rely on an attenuated laser pulse
for a single photon.

In the system studied by Law {\it et al.} \cite{law},
a single atom is placed inside a high Q optical cavity.
A pictorial illustration of the required energy level structure is
reproduced in the left panel of Fig. \ref{fig1}, where a three
level $\Lambda$-type atom with one excited state $|e\rangle$ and
two long lived ground states ($|g_1\rangle$ and $|g_2\rangle$)
are coupled respectively to a classical pump field (in solid line)
and the cavity field (in dashed line).
In the strong coupling limit $g\gg\gamma$ and $g\gg \kappa$ \cite{sc},
the dominant absorption-emission process consists of an atom
starting in $|g_1\rangle$, pumped into the excited state
$|e\rangle$, which then decays via the cavity into $|g_2\rangle$
\cite{law,kuhn}. Following the emission, an external laser
field driving the transition $|g_2\rangle\rightarrow |e\rangle$
can reset the atom to state $|g_1\rangle$ and prepare it for the
next photon emission. Such a single photon `gun' is expected to
reach a rep-rate $\sim \kappa$, which is typically several MHz.
An alternative approach based
on adiabatic passage for a deterministic or
`push button' single photon source was considered in \cite{kuhn}.
In the above, $g=d\cdot E_{0}/\hbar$ is the dipole coupling
between the atom and a single cavity photon
(of angular frequency $\omega$) field
$E_{0}=\sqrt{2\pi \hbar \omega /V}$
confined in a mode volume $V$. $d$ is the electric dipole
matrix element, and
$\gamma$ ($\kappa$) is the excited atom (cavity) decay rate.

A simple extension of the three-level atom to a four-level one as
illustrated in the right panel of Fig. \ref{fig1} consists of our
model. The excited state $|e\rangle$ is now resonantly coupled to
both states $|g_+\rangle$ and $|g_-\rangle$ through the left and
right circular polarized cavity photon field \cite{2g}. Following
Law {\it et al.} \cite{law}, the coherent part (in the rotating
wave approximation) of our model Hamiltonian can be expressed as
\begin{eqnarray}
H_0 &=&\hbar g(a_{L}\sigma_{e,g_-}+a^\dag_{L}\sigma_{g_-,e}
+a_{R}\sigma_{e,g_+}+a_{R}^\dag\sigma_{g_+,e})\nonumber\\
&&+{1\over 2}\hbar\Omega(t)(\sigma_{g_1,e}+\sigma_{e,g_1}),
\end{eqnarray}
where $\sigma_{\mu,\nu}(t=0)=|\mu\rangle\!\langle \nu|$
($\mu,\nu=g_1,e,g_-,g_+)$ are atomic projection operators.
$a_{\xi}$ and $a_{\xi}^\dag$ ($\xi=L,R$) are annihilation
and creation operators for the quantized cavity field.
$\Omega(t)$ denotes the coupling
between the atom and the external classical field.
Including the non-Hermitian dynamics due to both atomic spontaneous
decays and the cavity decay, the master equation of our system becomes
\begin{eqnarray}
{d\over dt}{\rho}&=&-\frac{i}{\hbar}[H_0,\rho]
+\kappa\sum_{\xi=L,R}(2a_{\xi}\rho a_{\xi}^\dag
-a_{\xi}^\dag a_{\xi}\rho-\rho a_{\xi}^\dag a_{\xi}) \nonumber\\
&& +\sum_{\mu=g_1,g_-,g_+}
\frac{\gamma \beta_{\mu}}{2}(2\sigma_{\mu,e}\rho\sigma_{e,\mu}
-\sigma_{e,e}\rho-\rho\sigma_{e,e}),
\end{eqnarray}
where $\beta_\mu$ denotes the branching ratio of
the atomic decay to levels $|\mu\rangle$ and
$\beta_{g_1}+\beta_{g_-}+\beta_{g_+}$=1.
Similar to the Law protocol \cite{law}, the
system is prepared in state $|g_1\rangle$ with no photon
in the cavity, after the classical field $\Omega(t)$ is tuned
on for a period $T_{0}$, a single photon (with either a left
or a right circular polarization) is generated in the cavity,
which immediately transmits outside the cavity in the bad
cavity limit of
\begin{equation}
\kappa\gg g^{2}/\kappa\gg\gamma.
\label{cp}
\end{equation}
The probability of detecting a photon with a given
polarization is easily computed according to
\begin{equation}
P_{\xi}(t)=2\kappa\int_{0}^{t}
\langle a_{\xi}^\dag(t')a_{\xi}(t')\rangle dt'.
\end{equation}

To gain more physical insight, we describe the dynamic evolution
of the system using the non-Hermitian effective Hamiltonian \cite{mp}
\begin{eqnarray}
H_{\rm eff}&&=H_0-i\hbar\kappa (a^\dag_{L}a_{L}+
a^\dag_{R}a_{R})-i\hbar\frac{\gamma}{2}\sigma_{e,e}.
\label{heff}
\end{eqnarray}
Using the combined basis of atomic internal state ($\mu=g_1,e,g_-,g_+$)
and the Fock basis of cavity photons $|\mu,n_L,n_R>$,
we can analyze the dynamics of the photon emission process.
Limited to $0$ and $1$ photon numbers $n_{L/R}=a_{\xi}^\dag a_{\xi}$,
the pure state wave function from Eq. (\ref{heff})
\begin{eqnarray}
|\psi(t)\rangle &&=a_{g_1}|g_1,0,0\rangle+a_{e}|e,0,0\rangle\nonumber\\
&&+a_{g_-}|g_-,1,0\rangle+a_{g_+}|g_+,0,1\rangle,
\end{eqnarray}
obeys the conditional dynamics described by
$i\hbar |\dot{\psi}\rangle=H_{\rm eff}|\psi\rangle$.
Explicitly, we find
\begin{eqnarray}
i\dot{a_{g_1}}(t) &&={1\over 2}\Omega(t)a_{e},\nonumber\\
i\dot{a_{e}}(t) &&={1\over 2}\Omega(t)a_{g_1}+ga_{g_-}+ga_{g_+}-i\frac{\gamma}{2}a_{e},\nonumber\\
i\dot{a_{g_-}}(t) &&=ga_{e}-i\kappa a_{g_-},\nonumber\\
i\dot{a_{g_+}}(t) &&=ga_{e}-i\kappa a_{g_+}.
\label{meq}
\end{eqnarray}
When the classical pump field satisfies the condition of
$\Omega(t)\ll g^{2}/\kappa$, the approximation
solution to Eq. (\ref{meq}) becomes
\begin{eqnarray}
a_{g_1}(t)&&\approx
\exp\left[-\frac{1}{2(4g^{2}/\kappa+\gamma)}\int_{0}^{t}\Omega^{2}(t')dt'\right],\nonumber\\
a_{e}(t) &&\approx-i\frac{\Omega(t)}{4g^{2}/\kappa+\gamma}\,a_{g_1}(t),\nonumber\\
a_{g_{\mp}}(t) &&\approx-i\frac{g}{\kappa}a_{e}(t),
\end{eqnarray}
given the initial condition of
$a_{g_1}(0)=1$ and $a_{e/g_-/g_+}(0)=0$.
Clearly, the left and right polarized modes are equally populated
if their couplings to the cavity mode are identical.
The conditional state of the system then becomes \cite{van,guo}
\begin{equation}
\frac{1}{\sqrt{2}}(|g_-\rangle|n_L=1,0\rangle
+|g_+\rangle|0,n_R=1\rangle),
\end{equation}
an atom-photon entangled state.

\begin{figure}
\includegraphics[width=2.75in]{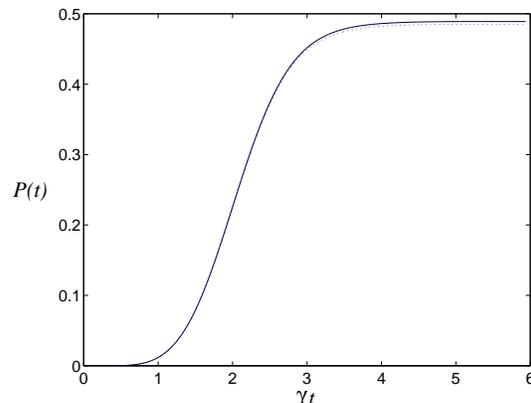}
\caption{The conditional probability for the emission of a cavity photon.}
\label{fig2}
\end{figure}

\begin{figure}
\includegraphics[width=2.75in]{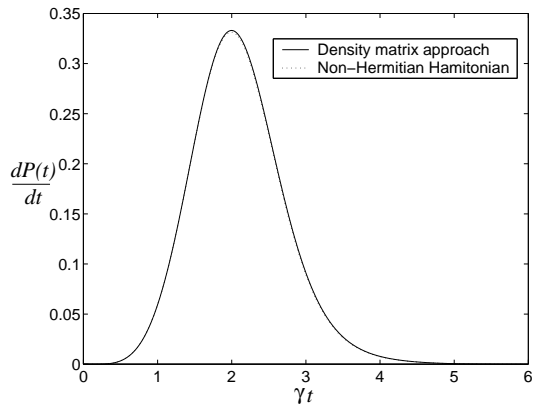}
\caption{Cavity emission rate}
\label{fig3}
\end{figure}

We have performed detailed numerical simulations for a
classical pump field of the form
\begin{equation}
\Omega(t)=\Omega_{0}\sin^{2}\left(\frac{\pi t}{T_{0}}\right), \qquad 0\leq t\leq T_{0},
\end{equation}
with $(g,\kappa,\gamma,\Omega_{0})=(2\pi)(45,45,4.5,45)$
(MHz) and $T_{0}=6/\gamma$=210 (ns).
We present selected results in Figs. \ref{fig2} and
\ref{fig3}. Clearly, our model works in exactly the same
way as the original Law protocol \cite{law}. The only difference
being now that the confirmed emission of a cavity photon
is accompanied with atom-photon entanglement. With the above
parameters, we find that the conditional probability for
a left or right polarization photon
rapidly increases to about $49\%$. For comparison,
we have solved both the conditional wave function dynamics
as well the the complete master equation dynamics.
The non-Hermitian Hamiltonian dynamics gives a slightly
lower value for final photon emission probability because
it excludes repeated spontaneous decays.
When the atom in $|e\rangle$ decays to $|g_1\rangle$,
it may be re-excited by the classical field before
emitting the photon into the right cavity mode.
Such an event should be excluded in order to have have
a final state with atom cavity coherence,
yet it is included in the master equation solution.

We note that the bad cavity limit, or the operating condition
as specified by the Eq. (\ref{cp}), in fact corresponds to the
cavity QED system {\it NOT} in the strong coupling limit.
 Thus the cavity photon
decays immediately once created, and this allows for an adiabatic
description by eliminating the atomic dynamics in the cavity. The condition
of $g^2\gg \kappa\gamma$ is in fact the same requirement of
a large cooperativity parameter $(C\propto g^2/\kappa\gamma)$
or a small critical atom number $(n_c\propto \kappa\gamma/g^2$)
as in the strong coupling limit. It turns out that this parameter
is an important characterization for the fidelity of several
important quantum computing protocols of atomic qubits inside high
Q cavities \cite{yi}. We now further illuminate this in terms
of the basic element of quantum information exchange between a
cavity and an atomic qubit.

We consider a three level $\Lambda$-type coupling scheme as
in the left panel of Fig. \ref{fig1}.
When the classical field $\Omega(t)$ is Raman resonant with respect to
the cavity photon (assuming a perfect compensation for ac Stark shifts \cite{blatt}),
while strongly off-resonant with respect to the atomic transition
$|e\rangle\leftrightarrow|g_2\rangle$,
the two states $|g_1,0\rangle$ and $|g_2,1\rangle$ are
effectively coupled directly
through a Rabi frequency $\Omega_{\rm eff}$ and an
effective atomic decay rate $\gamma_{\rm eff}$ given by,
\begin{eqnarray}
\Omega_{\rm eff} &&={1\over 2}{\Omega g\over \Delta},\\
\gamma_{\rm eff} &&={1\over 4}{\Omega^2 \over \Delta^2}\gamma,
\end{eqnarray}
where $\Delta=\omega_L-(\omega_e-\omega_1)$ is the pump field detuning.

To expect coherent dynamics for state mapping \cite{zo1} according to
\begin{eqnarray}
(\alpha|g_1\rangle+\beta|g_2\rangle)\otimes |0\rangle\to
|g_2\rangle\otimes (\alpha|1\rangle+\beta|0\rangle),
\label{sm}
\end{eqnarray}
one requires $\Omega_{\rm eff}\gg \gamma_{\rm eff}$ and
$\Omega_{\rm eff}\gg \kappa$,
which reduces to
\begin{eqnarray}
{\kappa\over g}\ll {\Omega\over \Delta}&&\ll {g\over\gamma},
\end{eqnarray}
thus $g/\gamma\gg \kappa/g$, or $g^2\gg \kappa\gamma$ \cite{chris}.

\begin{figure}
\includegraphics[width=3.in]{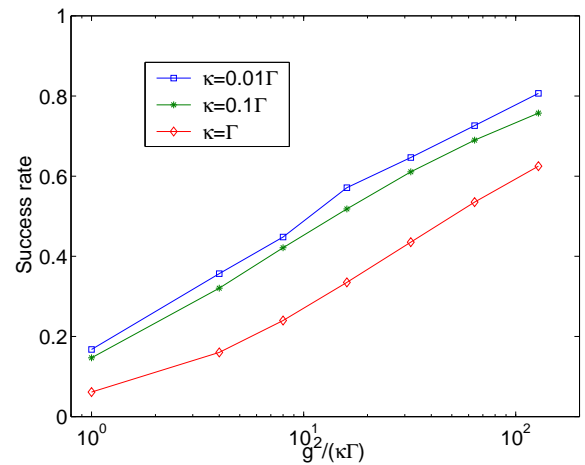}\\
\caption{The success rate of mapping the atomic state
$\alpha|g_1\rangle+\beta|g_2\rangle$ into the
cavity state $\alpha|1\rangle_C+\beta|0\rangle_C$.
The worst case scenario of $\alpha=1$ and $\beta=0$ is considered here.
States with non-zero $\beta$s generally
lead to proportionally larger success rates.
We have used $\gamma=(2\pi) 20$ (MHz) and $\Delta=50\gamma$.}
\label{fig4}
\end{figure}

We have performed extensive numerical simulations to check this
understanding. First for the Raman scheme and a constant $\Omega$,
we define {\it success rate} (Fig. \ref{fig4}) \cite{law}
as the conditional probability
for atom to end up in state $|g_2\rangle$ (from initially in state $|g_1\rangle$),
i.e. conditioned on the system to experience no spontaneous emission from either
the atom or the cavity.
The numerical results are given in Fig. \ref{fig4}, which shows
a weak dependence on the classical field detuning $\Delta$. Typically,
we find the optimal condition corresponds to $\Omega/\Delta\sim 0.2$-$0.75$.

\begin{figure}
\includegraphics[width=3.in]{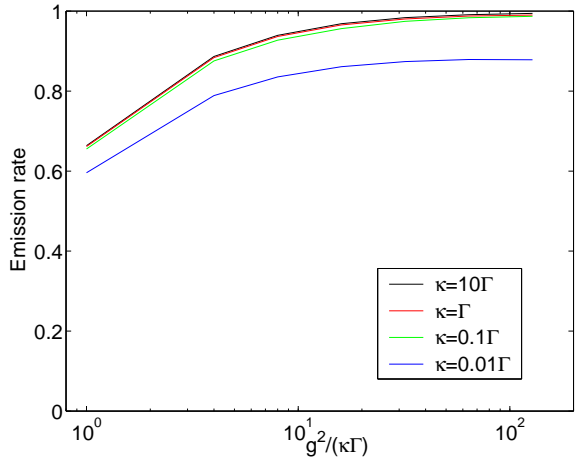}\\
\caption{The optimal photon-gun-quality
for $\gamma=(2\pi) 20$ (MHz) and $\Delta=50\gamma$.}
\label{fig5}
\end{figure}

The second figure of merit applies to the operation of the
atom + cavity system as a photon gun,
the aim of our proposed model. In this case,
the probability of emission, or the {\it emission rate} into
the cavity mode is used. In a sense,
it measures the photon-gun-quality. The results from
our numerical survey are illustrated in Fig. \ref{fig5}.
It is interesting to note that the results {\it DO}
depend on the detuning, essentially reflecting an
unbalanced choice of $\kappa$ with $\gamma_{\rm eff}$.
We also note that together with the probability of
the atomic spontaneous emission, the two add to unity
in the long time limit.

Building on several current experiments,
it seems possible to achieve $g^2/(\gamma\kappa)=30$ \cite{ch1,newkb,rempe},
a condition for very efficient photon gun according to our calculations;
in the strong coupling limit, this also becomes a promising
parameter regime for converting an atomic qubit
into a flying qubit (of 0 and 1 photons).

\begin{figure}
\includegraphics[width=3.in]{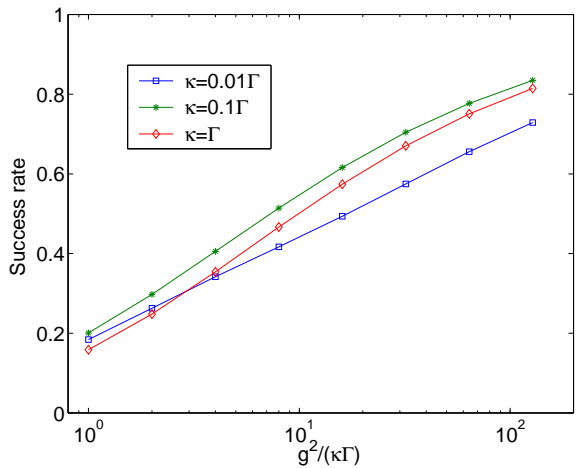}\\
\caption{The same as in Fig. \ref{fig4} but
with the adiabatic passage protocol. The results are
optimized and are observed to be less sensitive
on the ratio of $\kappa/\gamma$.
$\gamma=(2\pi) 20$ (MHz).}
\label{fig6}
\end{figure}

\begin{figure}
\includegraphics[width=3.in]{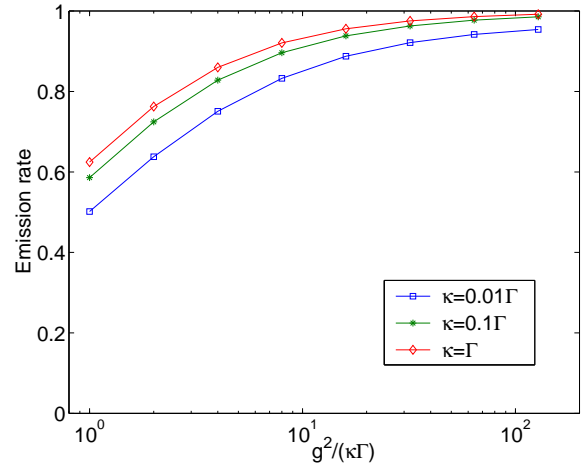}\\
\caption{The same as in Fig. \ref{fig5} but
with the adiabatic passage protocol. The results are
optimized and
for $\gamma=(2\pi) 20$ (MHz).}
\label{fig7}
\end{figure}

For comparison, we have also compared the state mapping Eq. (\ref{sm})
with the counter-intuitive pulse sequence for adiabatic passage \cite{zo1,hr1,lange,dark1}.
The best numerical results are shown in Figs. \ref{fig6} and \ref{fig7},
respectively. We note that in this case
the numerical survey is rather cumbersome as we are looking at a
3-dimensional ($\Omega$, $\Delta$, and $\delta$)
optimization search for each data point.
The Raman differential
detuning is defined as
$\delta=(\omega_L-\omega_C)-(\omega_{2}-\omega_1)$.

In summary, we have proposed a simple and efficient
implementation for a deterministic generation of atom-photon
entanglement. Our arrangement can be directly adopted
for distributing entanglement shared between any two parties
as the single photon can be propagated to reach a distant party.
Successfully realization of the controlled interactions between a single
trapped atom and a cavity photon will represent an important
milestone in Quantum Information Physics.

We acknowledge the contribution of X. H. Su during the early stage
of this work. Our research is supported by a grant from
the National Security Agency (NSA), Advanced Research and
Development Activity (ARDA), and the Defense
Advanced Research Projects Agency (DARPA) under Army Research Office
(ARO) Contract No. DAAD19-01-1-0667, and by NSF.


\begin{thebibliography}{}
\bibitem{chris}
Dr. Monroe (see also \cite{steane})
suggested the following order of magnitude estimates
to understand this requirement. The fidelity of state mapping is approximately given by
${\cal F}=1-p_{\rm atom}-p_{\rm cavity}$ where $p_{\rm atom}$ and $p_{\rm cavity}$
are respectively the rate for atomic spontaneous or cavity decay
given by $\gamma_{\rm eff}/\Omega_{\rm eff}\sim \Omega\gamma/(\Delta g)$
and $\kappa /\Omega_{\rm eff}\sim \Delta\kappa/(\Omega g)$. The maximal
for the fieldity therefore is approximatly given by
${\cal F}\le 1-2\sqrt{p_{\rm atom}p_{\rm cavity}}=1-2\sqrt{\kappa\gamma/g^2}$,
and the maximal occurs roughly when $p_{\rm atom}\sim p_{\rm cavity}$,
or $\gamma_{\rm eff}\sim \kappa$,
not within the bad cavity limit.
\end{thebibliography}
\end{document}